\documentclass[sigconf]{acmart}

\AtBeginDocument{%
  }

\acmConference[CIKM 2024]{Decoding Style: Efficient Fine-Tuning of LLMs for Image-Guided Outfit Recommendation with Preference Feedback}{October 25,
  2024}{Boise, ID}
\usepackage{booktabs}
\begin{document}

\title{Decoding Style: Efficient Fine-Tuning of LLMs \\ for Image-Guided Outfit Recommendation with Preference Feedback}


\author{Najmeh Forouzandehmehr}
\email{najmeh.forouzandehmehr@walmart.com}
\affiliation{%
  \institution{Walmart Global Tech}
  \city{Sunnyvale}
  \state{California}
  \country{USA}
}

\author{Nima Farrokhsiar}
\email{nima.farrokhsiar@walmart.com}
\affiliation{%
  \institution{Walmart Global Tech}
  \city{Sunnyvale}
  \state{California}
  \country{USA}
}

\author{Ramin Giahi}
\email{ramin.giahi@walmart.com}
\affiliation{%
  \institution{Walmart Global Tech}
  \city{Sunnyvale}
  \state{California}
  \country{USA}
}

\author{Evren Korpeoglu}
\email{ekorpeoglu@walmart.com}
\affiliation{%
  \institution{Walmart Global Tech}
  \city{Sunnyvale}
  \state{California}
  \country{USA}
}

\author{Kannan Achan}
\email{kannan.achan@walmart.com}
\affiliation{%
  \institution{Walmart Global Tech}
  \city{Sunnyvale}
  \state{California}
  \country{USA}
}


\begin{abstract}
Personalized outfit recommendation remains a complex challenge, demanding both fashion compatibility understanding and trend awareness. This paper presents a novel framework that harnesses the expressive power of large language models (LLMs) for this task, mitigating their "black box" and static nature through fine-tuning and direct feedback integration. We bridge the item visual-textual gap in items descriptions by employing image captioning with a Multimodal Large Language Model (MLLM). This enables the LLM to extract style and color characteristics from human-curated fashion images, forming the basis for personalized recommendations. The LLM is efficiently fine-tuned on the open-source Polyvore dataset of curated fashion images, optimizing its ability to recommend stylish outfits. A direct preference mechanism using negative examples is employed to enhance the LLM's decision-making process. This creates a self-enhancing AI feedback loop that continuously refines recommendations in line with seasonal fashion trends. Our framework is evaluated on the Polyvore dataset, demonstrating its effectiveness in two key tasks: fill-in-the-blank, and complementary item retrieval. These evaluations underline the framework's ability to generate stylish, trend-aligned outfit suggestions, continuously improving through direct feedback. The evaluation results demonstrated that our proposed framework significantly outperforms the base LLM, creating more cohesive outfits. The improved performance in these tasks underscores the proposed framework's potential to enhance the shopping experience with accurate suggestions, proving its effectiveness over the vanilla LLM based outfit generation.
\end{abstract}



\maketitle

\section{Introduction}

The central problem addressed in this research lies at the intersection of technology and fashion, two rapidly evolving fields. The challenge is to create an automated, personalized outfit recommendation system that not only understands fashion compatibility but is also sensitive to current trends and individual preferences. Given the proliferation of online shopping and the ever-increasing consumer demand for personalized experiences, the development of such a system holds significant practical relevance and economic potential. Personalized outfit recommendation has witnessed a revolution driven by advancements in artificial intelligence (AI) and natural language processing (NLP). Recent years have seen a surge of deep learning-based approaches tackling two key tasks: understanding fashion compatibility and capturing current trends. This literature review delves into existing research efforts along these lines, highlighting their strengths, limitations, and paving the way for our proposed framework.

\subsection{Graph Neural Networks (GNNs)}
By modeling product relationships as a graph, GNNs reason about outfit compatibility by considering global outfit coherence \citet{han2017learning}. This approach tackles outfit compatibility at multiple levels (coarse- grained and fine-grained categories), leading to richer information propagation and improved recommendation accuracy. Additionally, research by  \citet{li2020hierarchical} proposes a hierarchical graph that captures both user-outfit and outfit-item relationships. This approach unifies compatibility modeling and personalized recommendation. It leverages embedding propagation for effective information aggregation and user preference refinement, achieving superior performance. While GNN-based approaches excel at modeling garment relationships for outfit recommendation, they face several hurdles. Data sparsity, especially for uncommon items, can hinder effective learning. Large-scale recommendation systems might struggle
with GNNs' scalability demands. Additionally, their complex inner workings (such as message passing and encoder- decoder architecture) make interpretability challenging, limiting error correction and fine-tuning recommendations. New users and items encounter the "cold start" problem due to the lack of connections in the graph, leading to subpar suggestions.

\subsection{Transformer-based Approaches}
While GNNs offer valuable insights into garment relationships, recent advancements in Transformer-based architectures have opened new avenues for outfit generation. These models excel at capturing long-range dependencies and complex semantic relationships within text, making them highly suitable for understanding fashion descriptions and generating cohesive outfit suggestions. Transformer-based approaches like OutfitTransformer \citet{sarkar2022outfittransformer} leverage self-attention mechanisms to capture the intricate relationships between individual clothing items within an outfit. This holistic view enables them to generate outfits that are not only stylistically compatible but also consider global coherence and overall aesthetics.
Unlike GNNs, Transformers lack inherent scalability limitations and can effectively handle large datasets of fashion items and descriptions. This opens up their potential for application in large-scale recommendation systems. Transformers, while promising for their efficiency and interpretability, face limitations in capturing the crucial aspects of trend awareness and personalized context for effective outfit generation. Their focus solely on item embeddings limits their ability to consider broader factors like occasion, weather, user preferences, and current trends, potentially leading to unrealistic, impractical, or unoriginal suggestions. Additionally, inherent biases in the training data can be perpetuated, generating outfits that lack inclusivity or fail to reflect individual style needs. While the complex nature of self-attention mechanisms allows for understanding individual item relationships, it hinders transparency in the decision-making process, making it difficult to integrate user feedback and refine recommendations effectively. This limited grasp of context and trends stands in stark contrast to fine-tuned Large-scale Language Models (LLMs), which leverage their vast pre-trained knowledge to reason about these factors, generating more adaptable and contextually relevant outfit suggestions, albeit with potential trade-offs in computational cost and explainability.

LLMs, such as OpenAI's ChatGPT (2022) and GPT-4 (2023), have exhibited exceptional capabilities in various natural language processing (NLP) tasks. These LLMs usually comprise hundreds of billions of parameters and are mostly proprietary. This has led to the development of more accessible and cost-effective alternatives,
such as LLaMA \citet{touvron2023llama}. These alternatives utilize fine-tuning of open-source LLMs using task-specific data (e.g., ChatDoctor by \citet{li2023chatdoctor} ). However, full-model fine-tuning (FFT) poses significant computational and storage challenges, making it difficult for practical use. Before FFT became prevalent in LLMs, parameter-efficient fine-tuning (PEFT) \citet{houlsby2019parameterefficient} presented an appealing solution in the NLP domain, especially for pre-trained models like BERT \citet{devlin2019bert}. PEFT offers an efficient strategy for fine-tuning LLMs, enabling tuning of a limited set of external parameters instead of the entire model, while still achieving comparable or superior results \citet{weyssow2024exploring}. PEFT also provides an effective countermeasure against catastrophic forgetting. While fine-tuned LLMs offer immense potential for personalized outfit generation, their ability to perfectly align with human preferences and adapt to evolving trends remains a challenge.

Reinforcement Learning from Human Feedback (RLHF) methods, such as Proximal Policy Optimization (PPO) \cite{schulman2017proximal} and Direct Preference Optimization (DPO) \citet{rafailov2023direct}, offer promising avenues to enhance the alignment of these models with human expectations. LLMs can sometimes deviate from human expectations and produce outputs that, although grammatically correct, may lack style, coherence, or adherence to user intent. RLHF methods address this by incorporating human feedback into the
learning process. PPO, for example, trains a reward model based on human evaluations of generated outputs, guiding the LLM towards outputs that better align with human preferences. DPO simplifies the RLHF process by maximizing the likelihood of generating preferred outputs and minimizing dispreferred ones. This method requires a dataset of paired outputs, each with a prompt, two possible completions (preferred and dispreferred), and human annotations. By training the LLM to distinguish between these preferences, DPO can achieve comparable or superior results to PPO, with less computational power and potentially improved interpretability. Incorporating human feedback through RLHF allows LLMs to capture seasonal trends and user preferences for wearable outfits. The LLM learns to understand these nuances and generate recommendations that are stylish, practical, and contextually relevant by exposing the model to various outfits rated by human experts according to their suitability for different seasons, occasions, and styles. This research introduces a novel framework that addresses these challenges through:
\textbf{Fine-tuning LLMs:}
 We leverage fine-tuning techniques to empower LLMs with the ability to reason about fashion compatibility and current trends, while maintaining interpretability through direct user feedback integration.

\textbf{Bridging the Visual-Textual Gap:} To bridge the gap between textual descriptions and visual information inherent in fashion items, we employ multimodal language models (MLMs) capable of extracting style and color characteristics from human-curated fashion images. This enriched representation serves as the foundation for personalized recommendations.

\textbf{Continuous Self-Improvement:} To ensure ongoing relevance and adaptation, our framework implements a feedback loop that incorporates user preferences and dynamically incorporates seasonal trends into the LLM's decision-making process.
This paper rigorously evaluates the proposed framework on the Polyvore dataset \citet{han2017learning}, focusing on two key tasks: fill-in-the-blank and complementary item retrieval. We anticipate demonstrating the framework's effectiveness in generating stylish, trend-aligned outfit suggestions that continuously improve through direct feedback.
\section{Preliminary}
\subsection{Parameter-Efficient Training (PEFT)}
Parameter-Efficient Fine-Tuning (PEFT) enables adopting large pre-trained to specific task by training a small set of parameters. An important example of PEFT, Low-rank Adaptation (LoRA) \citet{hu2021lora} hypothesizes that the weight updates in pre-trained models have a low intrinsic rank during adaptation. For a pre-trained weight matrix,  $\mathbf{W} \in \mathbb{R}^{d \times k}$, it is updated with a low-rank decomposition $\mathbf{W} + \Delta \mathbf{W} = \mathbf{W} + \mathbf{A} \mathbf{B}$, where $\mathbf{A} \in \mathbb{R}^{d \times r}$ and $\mathbf{B} \in \mathbb{R}^{d \times k}$. The rank $r \ll \min(d, k)$. During training, $\mathbf{W}$ is frozen with no gradient updates, while $\mathbf{A}$ and  $\mathbf{B}$ are trainable. This is the reason why LoRA training is much more efficient than full fine-tuning. In the Transformer structure, LoRA only adapts the attention weights $(\mathbf{W}_q, \mathbf{W}_k, \mathbf{W}_v, \mathbf{W}_o)$ and freezes all other layers, including MLP and normalization layers. This manner is simple and parameter efficient. However, we empirically show that only low-rank adaptation in attention weights does not work for long context extension.

\subsection{ Reinforcement Learning from Human Feedback (RHLF)}
Reinforcement Learning from Human Feedback (RLHF), which aims to improve the natural language understanding capabilities of Language Learning Models (LLMs) by integrating human feedback during the training process. The RLHF method comprises three main stages:
\begin{enumerate}

\item  \textbf{Supervised Fine-Tuning (SFT)}: The RLHF
process typically starts with the fine-tuning of a pre-trained LLM through
supervised learning on high-quality data, targeting the downstream tasks of
interest such as dialogue, summarization, and more. This stage results in the creation of a model, denoted as $\pi^{SFT}$.

\item \textbf{Preference Sampling and Reward Learning}: In the second phase, the SFT model is given prompts x, which generate pairs of answers $(y_1, y_2) \sim \pi^{SFT}(y \mid x)$. These pairs are then shown to human labelers who express their preference for one answer. This preference is denoted as $y_{w} \succ y_{l} \mid x$ where $y_{w}$ and $y_{l}$ represent the preferred and less preferred completions amongst
$(y_1, y_2)$, respectively. A reward model $r(x,y)$ is then trained to rate the quality of the generated responses. 
\item \textbf{RL Fine-Tuning Phase}: In this phase, the SFT model (policy) is optimized under PPO reinforcement learning framework, with rewards calculated by Inverse Reinforcement Learning (IRL) using Kullback–Leibler ($KL$) divergence. The per-token probability distributions from the RL policy are compared to those from the initial model to compute a penalty on the difference between them. This penalty is designed as a scaled version of the $KL$ divergence between the sequences of distributions over tokens, denoted as $KL_r$. This $KL$ divergence term prevents the RL policy from deviating significantly from the initial pre-trained model with each training batch, ensuring the generation of reasonably coherent text snippets. In the third step, a PPO RL \citet{schulman2017proximal} formulation is used to fine-tune the model based on the reward function trained in the previous step.

\end{enumerate}
\subsection{Direct Preference Optimization (DPO)}
DPO has emerged recently as a viable alternative to RLHF for preference alignment, optimizing the policy model directly without needing to train a separate reward model and sample rewards through reinforcement learning \citet{rafailov2023direct}. It has shown comparable performances with RLHF in summarization and chatbot use cases on language models, and maintains strong performance in higher temperature sampling. At the same time, it avoids the unstable and brittle process of training models with RL.
DPO implicitly optimizes the same objective as existing RLHF algorithms (i.e., reward function with a KL-divergence term) discussed above. Specifically The DPO framework can be outlined in 2 steps:
\begin{enumerate}
\item For a given prompt $x$  generate completion pairs $(y_1, y_2)$ by sampling from the reference policy $\pi_{ref}(y \mid .)$.  These are then annotated with human preferences to form a comprehensive dataset $\mathcal{D} =\{ (\mathbf{x}^{i}, y^{i}_w, y^{i}_l)\}_{i=1}^{N}$

\item The next step involves refining the language model $\pi_{\theta}$ to reduce the DPO loss function $L_{DPO}$ considering the established  $\pi_{ref}$ , dataset $\mathcal{D}$, and the target ${\beta}$, where

\begin{align*}
L_{DPO} &= \mathbb{E}_{(x, y_{w}, y_{l})} \left[ \sigma \left( \beta \log \left(\frac{\pi_{\theta}(y_{w} \mid x)}{\pi_{ref}(y_{w} \mid x)}\right) \right. \right. \\
&\quad \left. \left. - \beta \log \left(\frac{\pi_{\theta}(y_{l} \mid x)}{\pi_{ref}(y_{l} \mid x)}\right) \right) \right]
\end{align*}
\end{enumerate}
$\beta$ is a parameter controlling the deviation from the base reference policy. Ideally, one prefers to capitalize on existing publicly available preference datasets to avoid the necessity of creating new samples and collating human preferences. Comprehensive details concerning the implementation and the hyperparameters are discussed in \citet{rafailov2023direct}.

\section{Methodology}
Our proposed model comprises four fundamental components:
MLLM image captioner, PEFT prompt generator, a pretrained large language model Mistral 7B, DPO prompt generator module. Figure \ref{fig:model} illustrates the overview of our framework. Our framework takes as input each outfit’s constituent item images. The components’ design and implementation
details are provided below:

\begin{figure}[h]
  \centering
  \includegraphics[width=\linewidth]{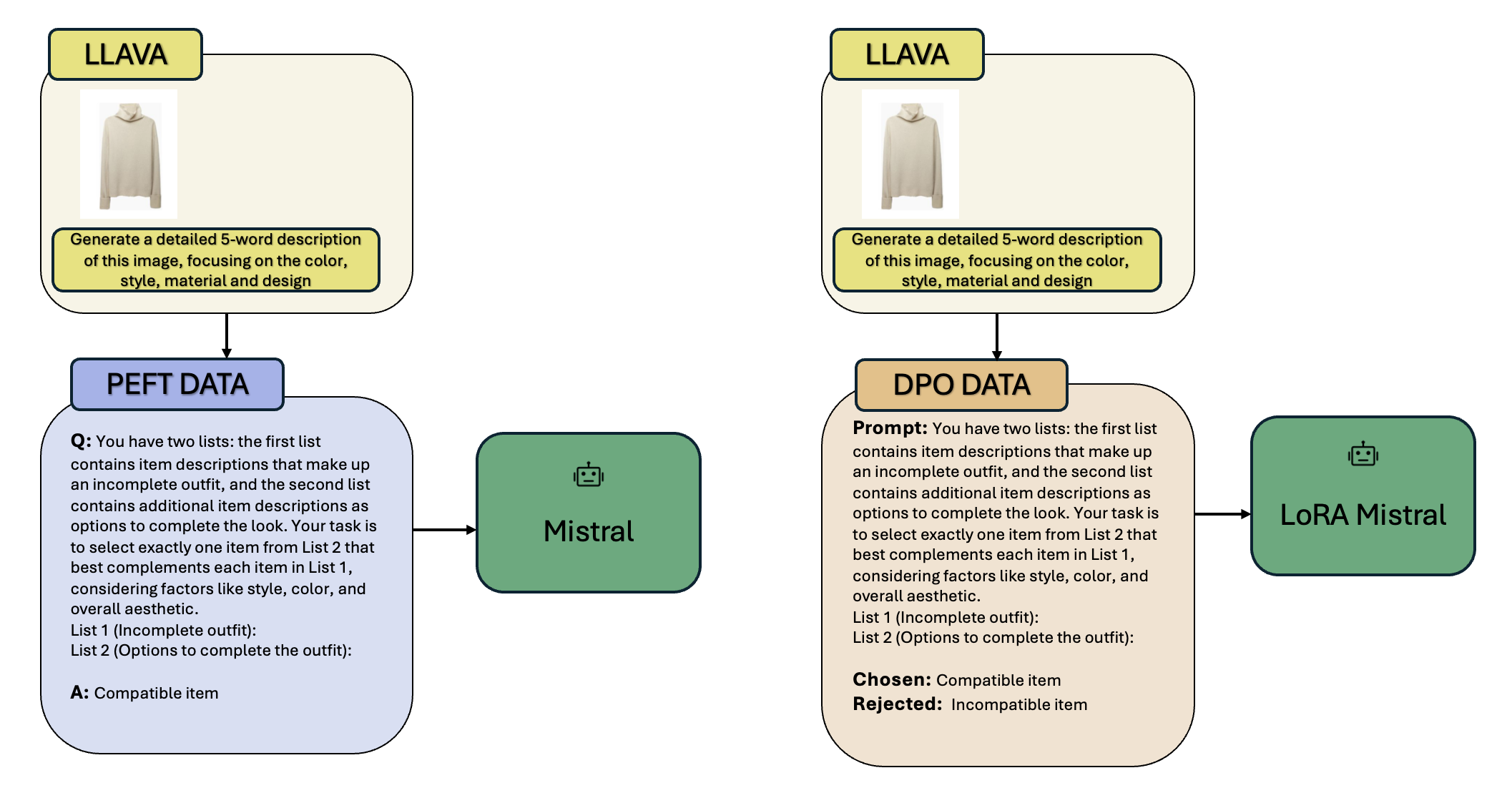}  
\caption{An overview of the proposed framework.}
\label{fig:model}
\end{figure}

\subsection{MLLM Image Captioner}
Multimodal Large Language Models (MLLMs) such as  LLaVA \citet{liu2023visual} are constructed by connecting a pre-trained vision
encoder with a LLM. The vision encoders are usually from CLIP \citet{radford2021learning} so that they can inherently extract semantically aligned visual features. The visual features are then adapted by a specialized light-weight module to map them into the hidden space of LLMs, so they can be jointly processed with the textual inputs by the LLM. Through multimodal training, the MLLMs learn to generate responses given the visual and textual inputs.
We used Llava as image captioning tool to provide detailed but short description of item images in each human curated outfits from Polyvore dataset, using the following prompt:
\begin{quote}
“Generate a detailed 5-word description of this image, focusing on the color, style, material and design”
\end{quote}
\subsection{PEFT Prompt Generator}
To align the pre-trained LLM model that recommends style cohesive outfits, we first perform supervised fine-tuning (SFT) on a pre-trained LM (Mistral 7B) on a small collection of labeled data. To collect the data, we generate specific prompts using items captions for each input $x$ and target output $y$ pair based on the downstream tasks ($CP$ and $FTTB$).

\subsubsection{Fill In The Blank (FITB) Task}
\begin{figure}[h]
\begin{center}
% Replace the placeholder with your actual image using the graphicx package
\includegraphics[width=9cm]{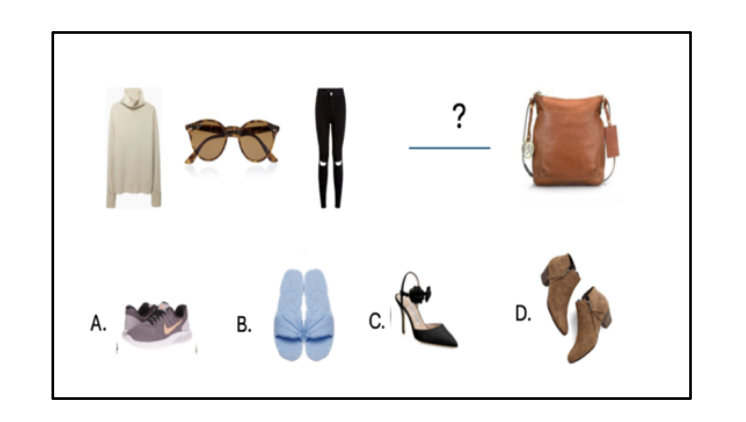}  
\caption{Example of Fill in The Blank (FITB) Task.}
\label{fig:ftib}
\end{center}
\end{figure}

In this task, a sequence of clothing and accessory items is provided, and LLM must choose an item from multiple choices compatible with other items to fill in the blank. This is a very practical scenario in real life, e.g., a user wants to choose a pair of shoes to match their pant and coat. An illustration of this task can be seen in Figure \ref{fig:ftib}.

We create a fill-in-the-blank dataset using curated outfits in the Polyvore set for this task, after image captioning each items a prompt from each training example constructed as follows:
\begin{quote}
“Human: You have two lists: the first list contains item descriptions that make up an incomplete outfit, and the second list contains additional item descriptions as options to complete the look. Your task is to select exactly one item from List 2 that best complements each item in List 1, considering factors like style, color, and overall aesthetic.
List 1 (Incomplete outfit)
List 2 (Options to complete the outfit) Assistant: (Correct Option)”
\end{quote}

\subsubsection{Outfit Compatibility Prediction (CP)}
\begin{figure}[h]
\begin{center}
% Replace the placeholder with your actual image using the graphicx package
\includegraphics[width=9cm]{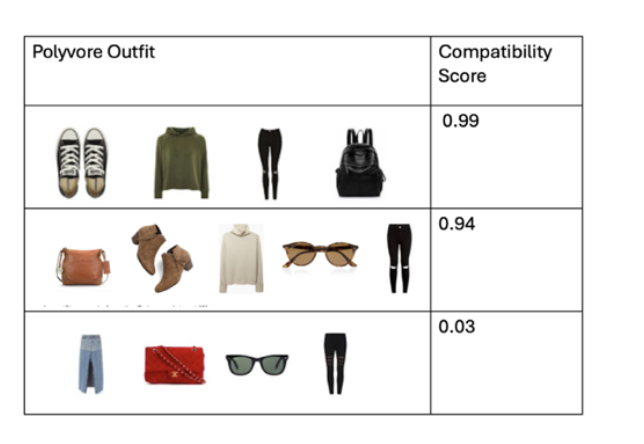}  
\caption{Example of Outfit Compatibility Prediction (CP) Task.}
\label{fig:cp}
\end{center}
\end{figure}

The task of compatibility prediction forecasts the compatibility of all items within an outfit. Examples of this task can be seen in Figure \ref{fig:cp}. To perform this task for each outfit from Polyvore disjoint training dataset, first we generate the item descriptions using image captioner step then create the prompt as follows:
\begin{quote}
“Human: As a fashion consultant, your task is to evaluate the overall style compatibility of a list of clothing item descriptions. You need to assign a single compatibility score between 0 and 1 for the entire list. A score of 1 indicates that the items are very compatible style-wise and can be combined to create a cohesive outfit. A score of 0 indicates that the items are not compatible at all.
List of clothing item descriptions (Complete Outfit)
Output: Compatibility score (0-1). Assistant: (Correct Score)”
\end{quote}
\subsection{Pretrained LLM}
For this research purpose we chose Mistral 7B \citet{jiang2023mistral} as the base LLM model, Mistral 7B developed by Mistral AI stands out as a powerful large language model (LLM) with 7 billion parameters. This decoder-based model utilizes a sliding window attention mechanism, allowing for efficient processing of long sequences with a theoretical attention span of 128K tokens. Additionally, its grouped query attention (GQA) enables faster inference and reduces cache size requirements. Notably, Mistral 7B surpasses Llama 2 13B on various demonstrating performance processing tasks.

\subsection{DPO Prompt generator}
To further enhance the alignment of Mistral LLM after LoRA fine tuning, we used DPO training as next training stage, the DPO training requires a dataset of prompts, each with two possible completions (preferred and dispreferred), we used the following adapted prompts for FITB and CP training tasks:
\subsubsection{Fill In The Blank (FITB)}
\begin{quote}
“You have two lists: the first list contains item descriptions that make up an incomplete outfit, and the second list contains additional item descriptions as options to complete the look. Your task is to select exactly one item from List 2 that best complements each item in List 1, considering factors like style, color, and overall aesthetic.
List 1 (Incomplete outfit)
List 2 (Options to complete the outfit), Chosen : (Correct Option),
Rejected: (Incorrect Option)”
\end{quote}

\subsubsection{Outfit Compatibility Prediction (CP)}

\begin{quote}
“As a fashion consultant, your task is to evaluate the overall style compatibility of a list of clothing item descriptions. You need to assign a single compatibility score between 0 and 1 for the entire list. A score of 1 indicates that the items are very compatible style-wise and can be combined to create a cohesive outfit. A score of 0 indicates that the items are not compatible at all. List of clothing item descriptions (Complete Outfit),
Chosen : (Correct score), Rejected: (1-Correct score)”
\end{quote}

The training process is conducted separately for both CP and FITB tasks, albeit in a similar manner. In the first stage, we apply PEFT to the pertinent Mistral-7b model on specific prompts, generating distinct labels for each task as previously demonstrated. During the second stage of training, we further refine the Mistral$\_$PEFT model using the DPO approach. For this stage, we make necessary adjustments to the training data to ensure its suitability for preference training as discussed in the previous stage.

\section{Evaluation}
For evaluation, we compare our proposed method with two baselines: plain LLM, and  PEFT LLM using LoRA on two different tasks:
\begin{enumerate}
    \item Outfit Compatibility Prediction (CP) task that predicts the compatibility of items in an outfit.
    \item Fill in the Blank (FITB) task that selects the most compatible item for an incomplete outfit given a set of candidate choices (e.g., 4 candidates).
    
The Polyvore Outfits dataset has two sets, the disjoint and non-disjoint sets. In the disjoint set, the training split items (and outfits) do not overlap with the validation and test splits.
\end{enumerate}
For this evaluation purpose we only focus on disjoint sets. The disjoint set comprises of 16995 training and 15154 test outfits. For the standard compatibility prediction and FITB tasks, we evaluate our model on the Polyvore Outfits dataset. We randomly selected 1000 outfits from the train set and generated PEFT and DPO prompts for each CP and FTIB training tasks. The training is done on one NVIDIA T4 GPU with a local batch size of 1 pair and gradient accumulation of 4 steps. We utilize AdamW optimizer with a learning rate of $2\times10^{-4}$ with 0. 3 warm-up steps and linear decay for PEFT training. We use same optimizer with a learning rate of $1\times10^{-8}$ with 0.1 ratio warm-up steps and linear decay for the DPO training. The DPO temperature parameter $\beta$ set be 0.1. The backbone model is trained for 3 epochs on our dataset.

\subsection{Outfit Compatibility Prediction (CP)}
The goal of this task is to measure the compatibility of an outfit. Our compatibility model predicts a score that indicates the compatibility of the overall outfit. We compare the performance with the baseline approaches in  Table~\ref{tab:cp} by using the standard metric AUC, which measures the area under the receiver operating characteristic curve. The purposed approach outperforms the other baseline approaches significantly.

\begin{table}[t]
 \caption{Comparison of our model with baselines on the CP task}
  \label{tab:cp}
  \begin{tabular}{ccl}
\toprule
\multicolumn{1}{c}{\bf Training Strategy}  &\multicolumn{1}{c}{\bf CP AUC} \\
\midrule
Plain LLM         &57.9\% \\
PEFT LLM (LoRA)            &62.27\% \\
PEFT DPO LLM            &81.03\%\\
\bottomrule
\end{tabular}
\end{table}

\subsection{FITB}
The goal of this task is to choose the best item from the entire database. For the FITB task we use accuracy. From Table~\ref{tab:fitb}, we observe that the purposed approach outperforms the other two baselines on the non-disjoint test dataset.

\begin{table}[t]
\caption{Comparison of our model with baselines on the FITB task}
\label{tab:fitb}
\begin{tabular}{ccl}
\toprule
\multicolumn{1}{c}{\bf Training Strategy}  &\multicolumn{1}{c}{\bf FITB Accuracy} \\
\midrule
Plain LLM         &29\% \\
PEFT LLM (LoRA)            &49\% \\
PEFT DPO LLM            &61\%\\
\bottomrule
\end{tabular}
\end{table}

\section{Conclusions}

This paper presents a novel framework for personalized outfit recommendation that leverages the power of large language models (LLMs) while mitigating their limitations through fine-tuning and direct feedback integration. We bridge the visual-textual gap by employing multimodal language models (MLMs) for image captioning, enabling the LLM to extract style and color features from fashion images. The future of this framework lies in expanding its capabilities by incorporating additional modalities like user context (location, occasion) to provide a richer understanding of preferences, while exploring advanced feedback mechanisms like click-through analysis to refine the learning process.

\bibliographystyle{ACM-Reference-Format}
\bibliography{ref}


\end{document}